\documentclass[a4paper,superscriptaddress,amsmath,amssymb,aps,twocolumn]{revtex4-2}
\usepackage{graphicx}
\usepackage{hyperref}
\usepackage{subfigure}

\hypersetup{
    colorlinks=true,
    linkcolor=blue,
    citecolor=blue,
    urlcolor=blue,
    hypertexnames=false
}

\begin{document}

\title{Scheduling the quantum implementation of the Trotter decomposition for Hamiltonian simulation}

\author{Augustin Ho}
\affiliation{Graduate School of Science and Technology, Keio University, 3-14-1 Hiyoshi, Kohoku, Yokohama, Kanagawa 223-8522, Japan}
\affiliation{\'Ecole des Mines de Saint-\'Etienne, 880 route de Mimet, Gardanne F-13541, France}

\author{Naoki Yamamoto}
\affiliation{Graduate School of Science and Technology, Keio University, 3-14-1 Hiyoshi, Kohoku, Yokohama, Kanagawa 223-8522, Japan}
\affiliation{Quantum Computing Center, Keio University, Hiyoshi 3-14-1, Kohoku, Yokohama, Kanagawa 223-8522, Japan}

\date{\today}

\begin{abstract}
    For the quantum computational task for Hamiltonian simulation, the arrangement in which sub-Hamiltonians appear in the Trotter decomposition may have a large influence on the precision, under a given circuit complexity.
    In this paper, we design several scheduling strategies by combining two methods; that is, the existing simultaneous diagonalization scheme leveraged in order to guarantee the performance in a relatively short circuit depth, with a method inspired from a Trotter-Suzuki decomposition of the second or the fourth order for a better accuracy.
    By conducting numerical simulations on various Hamiltonian instances, we confirm a systematic advantage over previous methods.
    Thus, our proposal may eliminate the need of dealing with several Hamiltonian-dependent scheduling strategies.
\end{abstract}

\maketitle

\section{Introduction}
\label{sec:intro}

In this work, we consider the quantum computing algorithm for time-independent Hamiltonian simulation, i.e. the task of calculating $\exp(-\mathrm{i} t \mathcal{H})$ for any $t \in \mathbb{R}$ and an $N$-dimensional Hamiltonian $\mathcal{H}$.
In practice, $N = 2^n$ for a quantum system of $n$ particles, which makes the complexities of classical procedures scale exponentially with respect to $n$.
In this spirit, it has been conjectured \cite{feynman-keynote} and later demonstrated that quantum computing offers a framework that bypasses this prohibitive cost.

The simplest form of Hamiltonian simulation algorithm relies on the first-order product formula called the Trotter decomposition \cite{quantum-simulators}.
More advanced techniques have been developed such as LCU \cite{truncated-taylor} and QSP \cite{qubitization}, but the product formula remains the most efficient in terms of resource requirements as it needs no ancilla qubit and no multi-qubit controlled gate \cite{survey}.
A critical problem in the product formula is how to arrange sub-Hamiltonians.
In the case of the first-order formula, this problem is described as follows.
Under the Pauli-access model, let $\mathcal{H} = \sum_{l=1}^{L} \alpha_l P_l$ be the Pauli decomposition of the Hamiltonian $\mathcal{H}$, where the $P_l$ are $n$-qubit Pauli operators.
Since $\mathcal{H}$ is Hermitian, all the $\alpha_l$ are real scalars.
Then, the Trotter decomposition for $\mathcal{H}$ is defined for any positive integer $r$ by the unitary operation
\begin{eqnarray}
    U_\mathcal{H}(t, r) &:=& \left(\prod_{l=L}^{1} \mathrm{exp}\left(-\mathrm{i} \frac{t}{r} \alpha_l P_l\right)\right)^r
    \label{eq:lie-trotter} \\
    &=& \underbrace{\mathrm{exp}\left(-\mathrm{i} \frac{t}{r} \alpha_L P_L\right) \times \cdots \times \mathrm{exp}\left(-\mathrm{i} \frac{t}{r} \alpha_1 P_1\right)}_{r \text{ times}}.
    \nonumber
\end{eqnarray}
The positive integer $r$ represents the number of Trotter steps which counts iterations of the time-ordered product sequence $\prod_{l=L}^{1} \mathrm{exp}\left(-\mathrm{i} \frac{t}{r} \alpha_l P_l\right)$ in $U_\mathcal{H}(t, r)$.
This product sequence is also referred to as the Trotter segment and is associated to an evolution duration of $t/r$.
The approximation-error of the first-order product formula is given by
\begin{equation}
    \varepsilon_\mathcal{H}(t, r) := \left\|\left(\prod_{l=L}^{1} \mathrm{exp}\left(-\mathrm{i} \frac{t}{r} \alpha_l P_l\right)\right)^r - \exp(-\mathrm{i} t \mathcal{H})\right\|,
    \label{eq:trotter-error}
\end{equation}
where $\|\cdot\|$ denotes the spectral norm.
The previous work \cite{trotter-theory} established a commutator scaling in the asymptotic regime $t/r \to 0$:
\begin{equation}
    \varepsilon_\mathcal{H}(t, r) = O\left(\frac{t^2}{r} \sum_{k=1}^{L} \sum_{l=1}^{L} \alpha_k \alpha_l \|[P_k, P_l]\|\right).
    \label{eq:commutator-scaling}
\end{equation}
However, this expression is not tight enough to comprehensively grasp the behavior of the Trotter error.
Especially, it does not reflect the fact that the way the Pauli operators are arranged in the product formula has an influence on $\varepsilon_\mathcal{H}(t, r)$.
In other words, we are claiming that how the implementation of the Trotter decomposition is scheduled may lead to a more accurate simulation.
The concept of scheduling the time-evolution refers to choosing the arrangement of the Pauli exponentiations in the circuit implementation of the time-ordered sequence (e.g. in Eq.~\eqref{eq:lie-trotter}, $P_1 \rightarrow P_2 \rightarrow \cdots \rightarrow P_L$).

This phenomenon has led researchers to investigate scheduling strategies in the attempt to offer a more effective error reduction.
We can for example mention the DepleteGroups strategy \cite{trotter-ordering} that has been designed to specifically target molecular Hamiltonian systems.
In addition to the Trotter error, scheduling choices also impact the circuit complexity, meaning the gate count and the circuit depth.
The latter aspect constituted the motivation for another scheduling approach featuring simultaneous diagonalization \cite{simult-diag}, as well as the so-called MaxCommuteTSP strategy \cite{circ-opt}.
However, all these techniques lead to limited improvements in terms of the error reduction or of the circuit complexity cut, both of which essentially depend on the Hamiltonian of interest.

In this work, we propose some new scheduling strategies that demonstrate performance improvements with respect to both the approximation-error and the circuit complexity.
In the literature, the circuit complexity is sometimes assimilated to the total number of sub-unitaries $Lr$, which corresponds to a high-level perspective.
In this paper, we consider its low-level definition, that is the complexity of the circuit described using a set of elementary quantum gates (e.g. $\{X, Y, Z, H, S, \text{CNOT}, R_Z(\theta)\}$).
In Section \ref{sec:method}, we will present our proposed methods.
Then, we will assess the performance of our techniques by testing them on various Hamiltonian systems in Section \ref{sec:result}.
Specifically, we compare the previous strategies \cite{trotter-ordering, circ-opt, simult-diag} with all the versions of our proposal.
Finally, we will discuss the limitations of our contribution as well as the opportunities it opens up in Section \ref{sec:conclusion}.

\section{Methods}
\label{sec:method}

Here we describe the proposed three methods.
The first one, roughly speaking, consists of combining the benefits of the techniques found in \cite{simult-diag} and \cite{circ-opt}.
Then, the second and third methods also feature this combination, except the sub-Hamiltonians are reorganized to make the structure of the product sequence follow the product formula of the second and fourth orders respectively.
This structural modification aims at reducing the algorithmic error (i.e. Trotter error) without changing the total number of sub-unitaries $Lr$.
Throughout this work, we consider the circuit depth to be the relevant metric for the circuit complexity.

\subsection{Simultaneous diagonalization with cell optimization}

\begin{figure*}
    \centering
    \includegraphics[width=\linewidth]{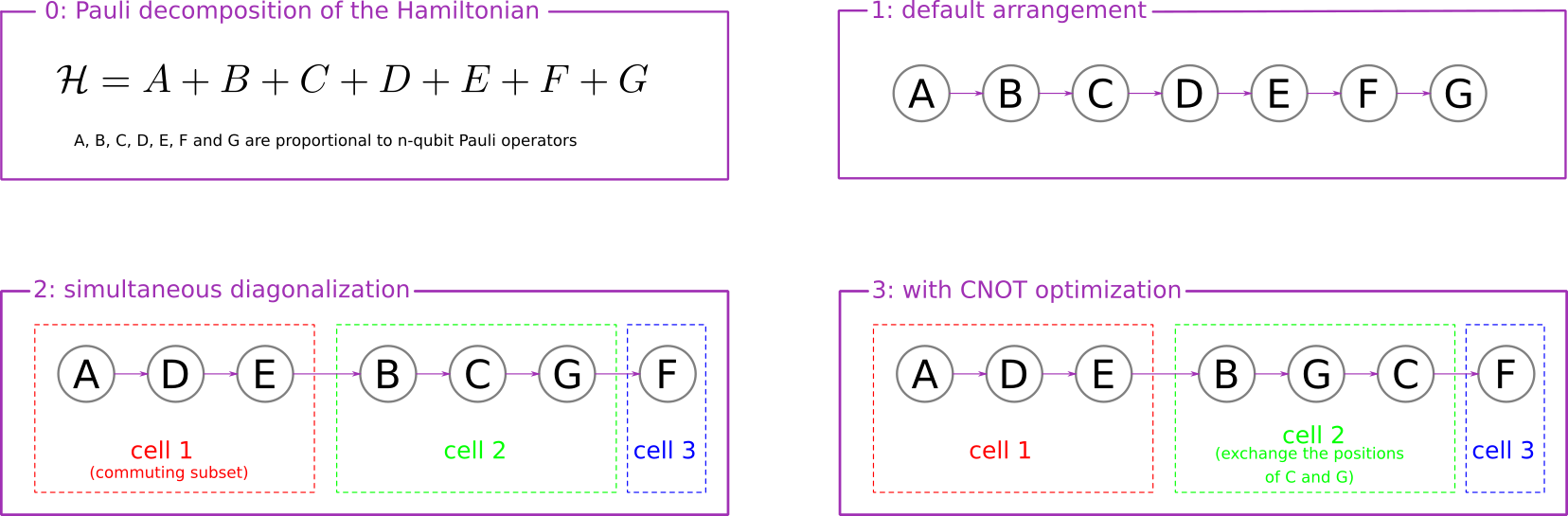}
    \caption{Construction of the simultaneous diagonalization with cell optimization strategy for a specific 7-term Hamiltonian. Cell 1, cell 2 and cell 3 do not commute and appear in the order as directly produced by the recursive-largest-first heuristic for the coloring graph problem.}
    \label{fig:simult-diag-cell-opt-proc}
\end{figure*}

Among the previous methods mentioned in Section \ref{sec:intro}, we found that the simultaneous diagonalization technique from \cite{simult-diag} often lead to the shortest circuit depth to achieve a target precision.
To further improve it, we make use of the same CNOT optimization procedure developed for the purpose of MaxCommuteTSP \cite{circ-opt}.
To be clear, this consists of arranging the Pauli operators in Eq.~\eqref{eq:lie-trotter}, within each commuting subset so that the CNOT count is minimized.
Overall, in our improved version of simultaneous diagonalization, we consider the commutation partition of the Hamiltonian $\mathcal{H}$, and within each cell (i.e. commuting subset), the optimization procedure to minimize the CNOT count is applied, hence the name ``Simultaneous diagonalization with cell optimization".
The construction of the strategy is explained in Fig.~\ref{fig:simult-diag-cell-opt-proc} for an example of $\mathcal{H}$.

To execute the simultaneous diagonalization, we first need to partition the Hamiltonian into commuting subsets.
This is done with the idea introduced in \cite{trotter-ordering}, by solving the graph coloring problem on a graph $G = (V, E)$, where the nodes represent the Pauli terms and the edges account for a lack of commutation between two terms.
Unlike the MaxCommuteTSP technique, the order in which the commuting subsets are simulated is not specified in \cite{trotter-ordering}; by default it follows the order produced by the graph coloring algorithm we leverage.
Here we take the recursive-largest-first heuristic, where the commuting subsets are ordered such that their sizes decrease.
Then, we compute the unitary operation corresponding to the simultaneous diagonalization, in the exact same way as presented in \cite{simult-diag}.
Finally, as we stated earlier in this section, the CNOT optimization procedure is drawn from the MaxCommuteTSP approach \cite{circ-opt}, in which the minimization of the CNOT count within each commuting subset is obtained by solving the Traveling Salesman Problem for another graph $G' = (V', E')$.
Here, each node corresponds to each Pauli term in the given commuting subset, and every edge between two terms $P_i$ and $P_j$ is weighted with, roughly speaking, the number of CNOT gates that appear in the circuit implementation of $\exp(-\mathrm{i} P_j) \times \exp(-\mathrm{i} P_i)$.

Note that the CNOT optimization only affects the arrangement within each commuting subset, and neither the commuting subsets themselves, nor the order in which they appear in the implementation change.
As a result, this modification in the simultaneous diagonalization strategy does not alter the Trotter error.
Thus, compared to the original simultaneous diagonalization procedure \cite{simult-diag}, our improved version is expected to yield shallower circuits for the same simulation accuracy.
However, we still have to address the error reduction in order to propose a scheduling strategy that guarantees an undeniable improvement over the previous methods.

\subsection{Cyclic-2}

To further improve the performance in terms of error reduction, we take the initiative to revisit the first-order product formula.
Indeed, the Trotter decomposition is defined by $r$ iterations of the sequence $\prod_{l=L}^{1} \mathrm{exp}\left(-\mathrm{i} \frac{t}{r} \alpha_l P_l\right)$, as given in Eq.~\eqref{eq:lie-trotter}.
The corresponding arrangement of the Pauli terms within each segment is denoted by $1 \rightarrow 2 \rightarrow \cdots \rightarrow L$.
Conventionally, it is the same sequence that is applied throughout the $r$ iterations.
We call such a scheduling mechanism ``static", as opposed to the ``dynamic" concept where the sequence may change from one iterate to another.

In our work specifically, we develop a dynamic strategy which we call ``cyclic-2".
Inspired by the resemblance between the first-order and the second-order product formulas, the idea is to alternatively apply the forward and the backward directions of a given arrangement.
In other words, we cyclically select from two sequences at each of the $r$ iterations, hence the name.
The given arrangement from which the forward and the backward directions are established is called the ``seed arrangement".
To put it mathematically, our cyclic-2 scheduling strategy implements the operation
\begin{equation}
    U_\mathcal{H}(t, r) = \left(\underbrace{\prod_{l=1}^{L} \mathrm{exp}\left(-\mathrm{i} \frac{t}{r} \alpha_l P_l\right)}_{\text{backward direction}} \, \underbrace{\prod_{l=L}^{1} \mathrm{exp}\left(-\mathrm{i} \frac{t}{r} \alpha_l P_l\right)}_{\text{forward direction}}\right)^{r/2},
    \label{eq:cyclic-2-even}
\end{equation}
when $r$ is an even number.
If $r$ is odd, the definition of the operation becomes
\begin{equation}
    \resizebox{\linewidth}{!}{$U_\mathcal{H}(t, r) = \displaystyle\prod_{l=L}^{1} \mathrm{exp}\left(-\mathrm{i} \frac{t}{r} \alpha_l P_l\right) \left(\displaystyle\prod_{l=1}^{L} \mathrm{exp}\left(-\mathrm{i} \frac{t}{r} \alpha_l P_l\right) \displaystyle\prod_{l=L}^{1} \mathrm{exp}\left(-\mathrm{i} \frac{t}{r} \alpha_l P_l\right)\right)^{^{(r-1)}\!/_2}$}.
    \label{eq:cyclic-2-odd}
\end{equation}
In Eqs.~\eqref{eq:cyclic-2-even} and \eqref{eq:cyclic-2-odd}, we assume the seed arrangement $1 \rightarrow 2 \cdots \rightarrow L$, but our cyclic-2 method can be abstractly defined for any seed arrangement.
So, the above equations can be rewritten in a more generic way by replacing $\alpha_l$ and $P_l$ with $\alpha_{\sigma(l)}$ and $P_{\sigma(l)}$ respectively, where $\sigma \in \mathfrak{S}_L$ is the permutation corresponding to the chosen seed arrangement.
Note that the implementation \eqref{eq:cyclic-2-even} is equivalent to the second-order product formula, except the number of Trotter segments is $r/2$ and thus the total number of sub-unitaries is maintained as $Lr$.
Therefore, our cyclic-2 strategy must achieve an approximation-error of the second order, that is to say $O\left(t^3/r^2\right)$ which converges faster than the first-order scaling presented in Eq.~\eqref{eq:commutator-scaling} as $r$ grows.
At this point though, one may argue that the cyclic-2 scheme does not bring any novelty from the second-order product formula \cite{sym-product}.
However, the number of Trotter segments being halved constitutes the key difference which makes the circuit complexity remain equivalent to the implementation of the first-order decomposition, thus enabling a fair comparison with the techniques developed in \cite{trotter-ordering, simult-diag, circ-opt}.

\begin{figure*}
    \centering
    \includegraphics[width=\linewidth]{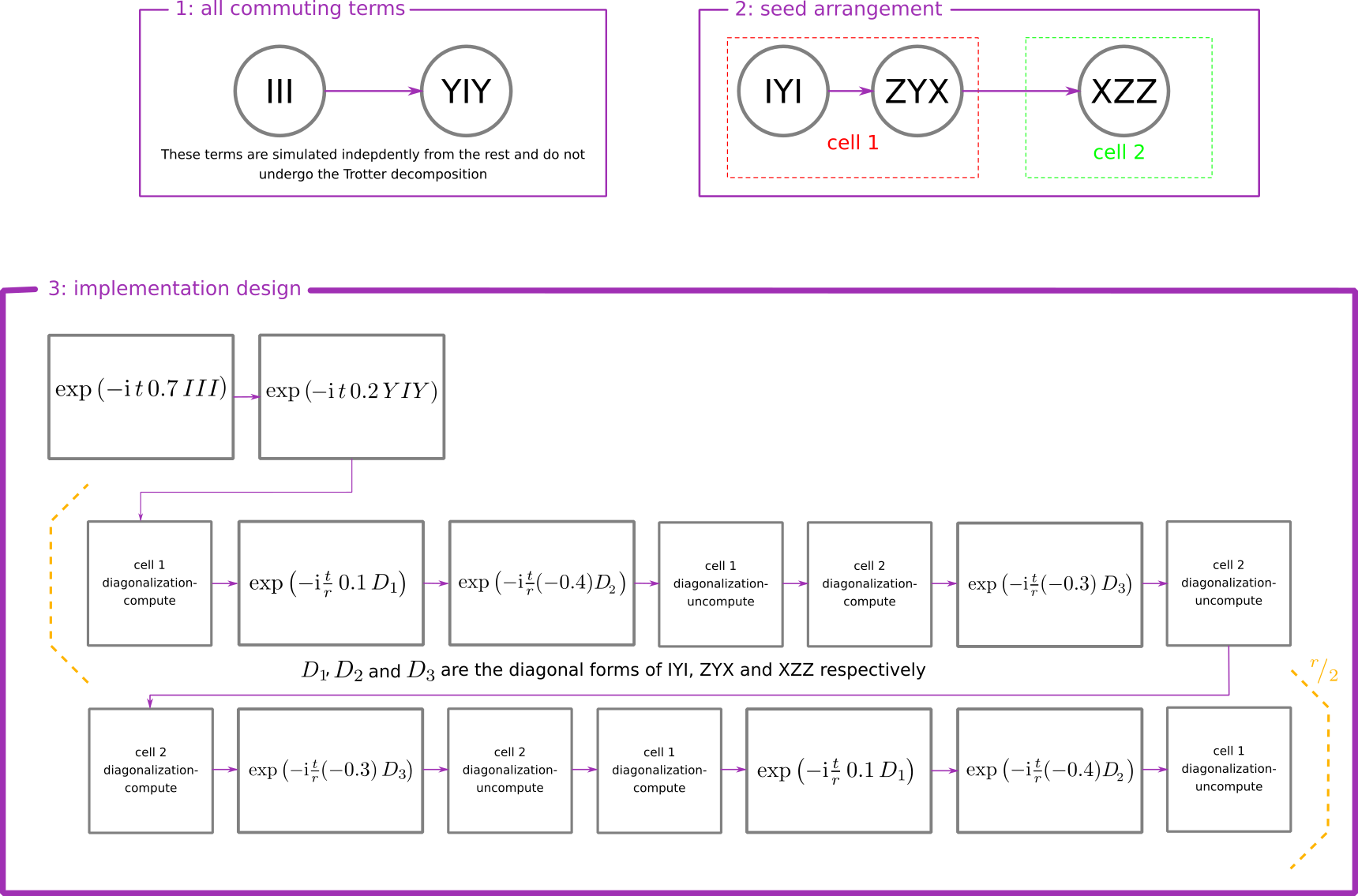}
    \caption{Description of our scheduling strategy.
    The first two operations are fully time-evolved at the beginning of the implementation. Then, the remaining Pauli terms undergo the Trotter decomposition with the number of Trotter steps $r$, according to the simultaneous diagonalization with cell optimization strategy in the forward direction (row in the middle of box 3) and the backward direction (lower row). The number of iterations of the product sequence is $r/2$. One may note that ``cell 2 diagonalization-uncompute" of the forward direction simplifies with ``cell 2 diagonalization-compute" of the backward direction. The same applies between ``cell 1 diagonalization-compute" of the forward direction and ``cell 1 diagonalzation-uncompute" of the backward direction.}
    \label{fig:top-level}
\end{figure*}

So far, we have introduced two scheduling techniques, namely the simultaneous diagonalization with cell optimization for the circuit complexity reduction, and cyclic-2 for the error convergence.
The idea now is to combine both of them so that we can benefit from the two advantages.
This is simply done by constructing cyclic-2 while considering our simultaneous diagonalization technique for the seed arrangement.
As suggested in \cite{trotter-ordering}, we exclude every Pauli operator that commutes with all other terms in the decomposition of the Hamiltonian beforehand.
Indeed, such operators do not contribute in the Trotter error whatsoever, hence they can be simulated independently from the rest.
Thus, no segmentation is required for their time-evolution operations, which implies a significant circuit depth reduction (especially if the number of completely commuting terms is large).
In our scheduling strategy, we arbitrarily choose to implement the completely commuting part upstream the time-evolution of the remaining Pauli terms in the quantum circuit.
For clarity, a concrete example picturing the entire procedure to implement our ``cyclic-2 under simultaneous diagonalization with cell optimization" technique is presented in Fig.~\ref{fig:top-level}, where the Hamiltonian is defined on 3 qubits and consists of 5 Pauli terms: $\mathcal{H} = 0.7 \, \texttt{III} + 0.1 \, \texttt{IYI} - 0.3 \, \texttt{XZZ} + 0.2 \, \texttt{YIY} - 0.4 \, \texttt{ZYX}$.

\subsection{Cyclic-10}

Recall that our cyclic-2 strategy reproduces the error scaling of the second-order product formula.
Given that, one may want to extend the approach to imitate a higher-order product formula to obtain an even faster convergence of the approximation-error, while keeping the total number of sub-unitaries equal to $Lr$.

Let us remind that higher-order product formulas, commonly referred to as ``Trotter-Suzuki decompositions" are only defined for an even order and are derived recursively from the second-order product formula.
In particular, the Trotter-Suzuki decomposition directly following the second-order product formula is the one of the fourth order, given by
\begin{equation}
    S_4(x) = \left(S_2(sx)\right)^2 \, S_2((1-4s)x) \, \left(S_2(sx)\right)^2,
    \label{eq:fourth-order}
\end{equation}
where $S_2(x) := \prod_{l=1}^{L} \mathrm{exp}(-\mathrm{i} \frac{x}{2} \alpha_l P_l) \prod_{l=L}^{1} \mathrm{exp}(-\mathrm{i} \frac{x}{2} \alpha_l P_l)$ is the second-order product formula for the Hamiltonian $\mathcal{H} = \sum_{l=1}^{L} \alpha_l P_l$, and $s := \frac{1}{4 - \sqrt[3]{4}}$ is such that $S_4\left(t/r\right)$ achieves an approximation-error of the fourth order (i.e. that scales as $O\left(t^5/r^4\right)$) \cite{higher-order}.
Having this formula, we explore in this work the extension of the cyclic-2 strategy to a scheme that corresponds to the fourth-order Trotter-Suzuki decomposition.
Accordingly to Eq.~\eqref{eq:fourth-order}, the implemented operation becomes
\begin{equation}
    \resizebox{\linewidth}{!}{$U_\mathcal{H}(t, r) = \left(\left(S_2\left(s \frac{2t}{r}\right)\right)^2 \, S_2\left((1-4s) \frac{2t}{r}\right) \, \left(S_2\left(s \frac{2t}{r}\right)\right)^2\right)^{r'}$},
    \label{eq:cyclic-10}
\end{equation}
where $r'$ is the number of Trotter segments such that the total number of product sequences in both the forward and backward directions is equal to the effective Trotter step count $r$.
Since $S_2$ induces two product sequences, the number of product sequences that appear in $S_4$ is 10, and thus $r' = \lfloor r/10\rfloor$.
If $r$ is not a multiple of 10, then $U_\mathcal{H}(t, r)$ is completed with respect to the successive product sequences that appear in the expression of $S_4(2t/r)$, equivalently to what we did for the cyclic-2 strategy in case of odd values for $r$.
Because it now takes 10 Trotter segments to exhaust all the product sequences in $S_4$ entirely, we call this dynamic scheduling scheme the ``cyclic-10" strategy.

\section{Numerical simulation results}
\label{sec:result}

We conduct numerical experiments for four Hamiltonian systems $\mathcal{H}$ for comparing the performance of cyclic-2 and cyclic-10 methods (in which the seed arrangement is given by simultaneous diagonalization with cell optimization) to the existing static strategies (that mainly consist of DepleteGroups, MaxCommuteTSP and traditional simultaneous diagonalization).
To assess the circuit complexity, we examine the CNOT count and the circuit depth against the Trotter step count $r$.
Then, we will analyze how the Trotter error $\varepsilon_\mathcal{H}(t, r)$ scales as a function of $r$.
Lastly, we will consider the Trotter error directly against the circuit depth as the relevant metric in the practical framework.

\subsection{Uniform combination of Pauli operators}

For the first application, let the Hamiltonian be the uniform combination of $n$-qubit Pauli operators:
\begin{equation}
    \mathcal{H} = \sum_{k=0}^{4^n-1} B_k,
    \label{eq:uniform-n-pauli}
\end{equation}
where $B_k$ are the $n$-qubit Pauli operators, i.e. $\ B_k := \bigotimes_{q=n-1}^{0} \pi_k^{(q)}$ with $\pi_k^{(q)} \in \{I; \sigma_X; \sigma_Y, \sigma_Z\}$).
In our case, we conducted the experiment with $n = 3$ qubits.
The simulation time is set to $t = 0.2$.

\begin{figure*}
    \centering
    \subfigure[]{\includegraphics[width=0.45\linewidth]{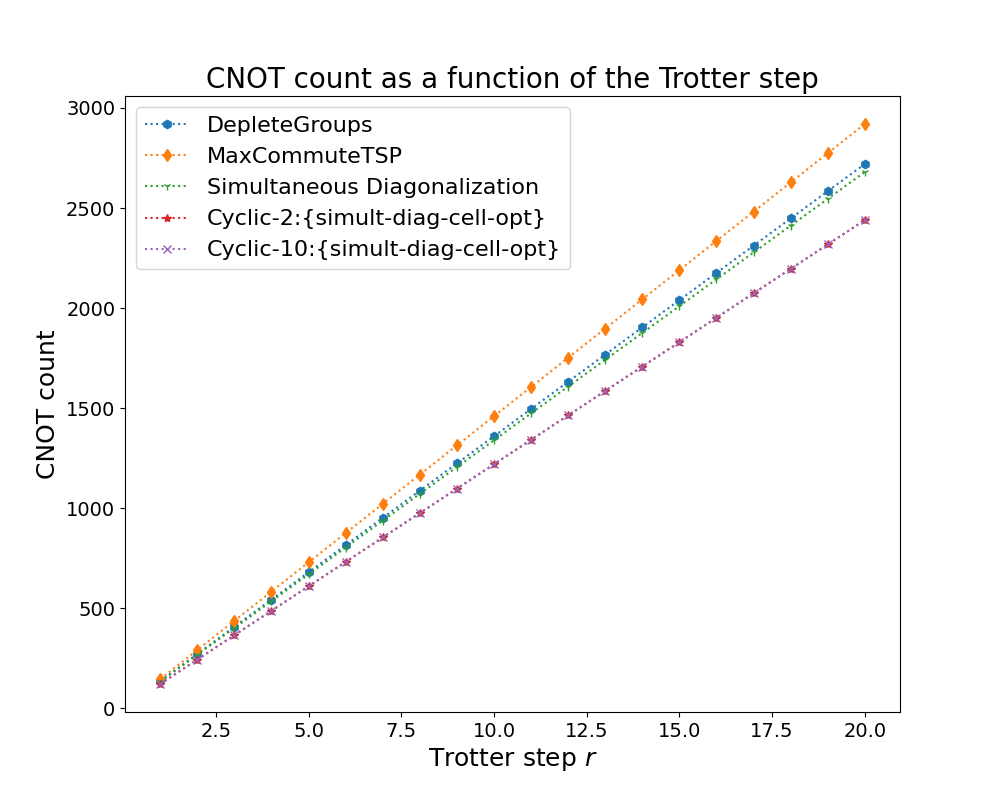} \label{fig:uniform-pauli-cnot}}
    \subfigure[]{\includegraphics[width=0.45\linewidth]{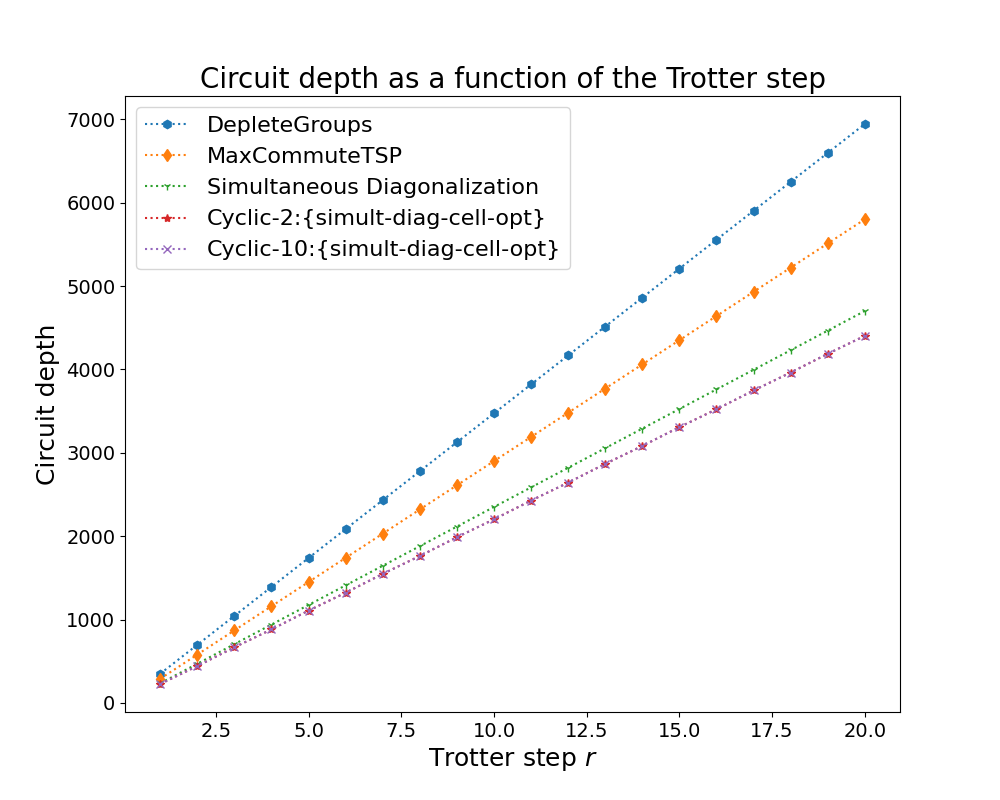} \label{fig:uniform-pauli-depth}}
    \subfigure[]{\includegraphics[width=0.45\linewidth]{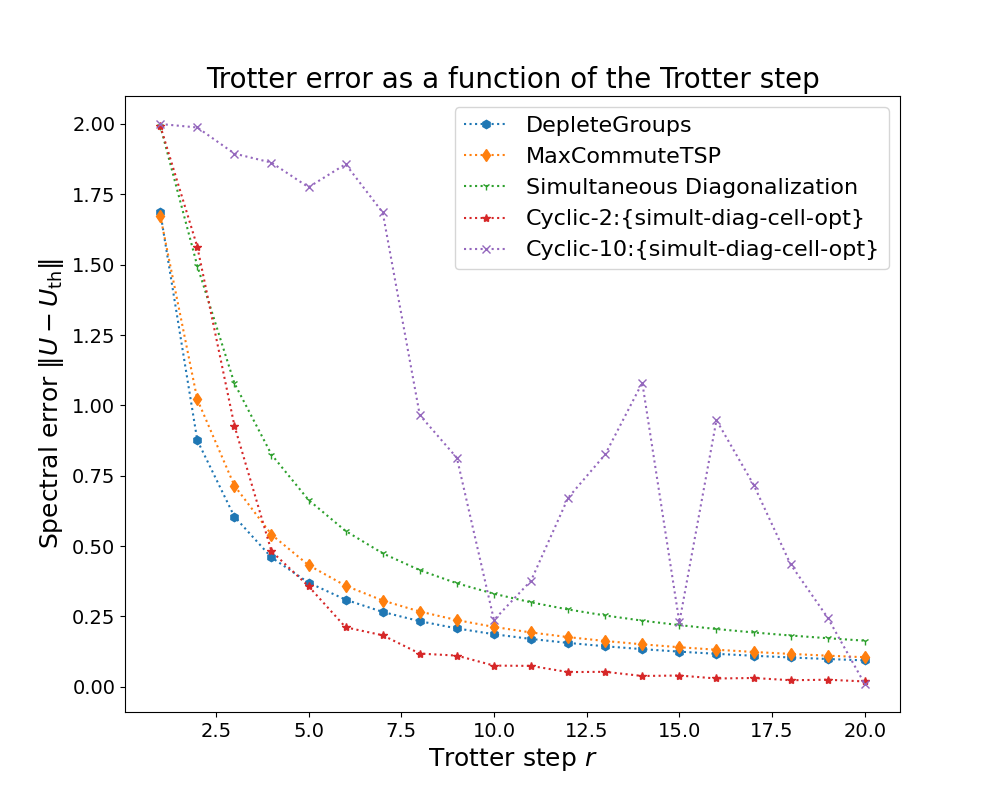} \label{fig:uniform-pauli-err}}
    \subfigure[]{\includegraphics[width=0.45\linewidth]{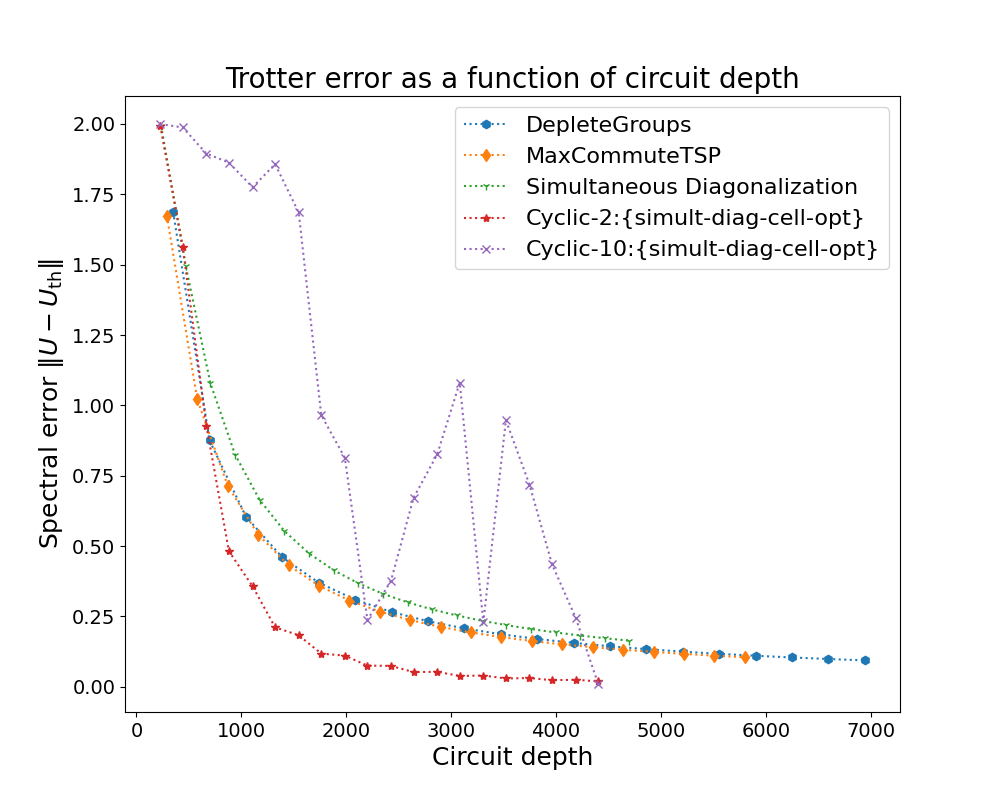} \label{fig:uniform-pauli-err-circ}}
    \caption{Performance comparison for the uniform combination of 3-qubit Pauli operators}
    \label{fig:results-uniform-pauli}
\end{figure*}

To begin with, Fig.~\ref{fig:uniform-pauli-cnot} gives the CNOT count as a function of the number of Trotter steps $r$.
As expected, both our strategies (cyclic-2 and cyclic-10) yield the smallest CNOT counts among the techniques we consider in this paper, which in particular provide a strict reduction from the case of bare simultaneous diagonalization.
This improvement is certainly explained by the use of the CNOT optimization in our methods.
As for the circuit depth (Fig.~\ref{fig:uniform-pauli-depth}), our strategies outperform again each of the previous approaches.
This may be thanks to simultaneous diagonalization, given shorter circuit depths obtained with its bare version compared to the performance of DepleteGroups and MaxCommuteTSP.

The results concerning the error reduction is shown in Fig.~\ref{fig:uniform-pauli-err}.
The fastest error convergence is obtained with cyclic-2.
The clear improvement over previous works is consistent with what we discussed in Section \ref{sec:method}, that is, cyclic-2 achieves an asymptotic error scaling of the second order.
However, the results regarding our approach using the cyclic-10 scheme, which is supposed to achieve an even better error scaling, features a negatively surprising behavior of the Trotter error as a function of the number of Trotter steps.
In fact, we particularly observe a detrimental error reduction for Trotter step counts that are not multiple of 10.
This could have been expected to some extent because cyclic-10 implements the fourth-order product formula only for numbers of Trotter steps that are multiples of the product sequence length which is 10 (Eq.~\eqref{eq:fourth-order}).
On the other hand, we do not see any equivalent artifact with cyclic-2 when $r$ does not correspond to a multiple of 2.
For this reason, we deem this robustness unique to cyclic-2, but the theoretical investigation is left for a future work.

Let us now put the results altogether and look at the error scaling against the circuit depth for practical relevance (Fig.~\ref{fig:uniform-pauli-err-circ}).
There is a unquestionable improvement of the performance by using cyclic-2 under simultaneous diagonalization with cell optimization instead of any other scheduling technique.
In the end, the use of cyclic-10 may lead to the practical issue that the fourth-order error scaling is only achieved against a Trotter step count that grows 10 times more than in the first-order product formula.
Thus, it would require a circuit depth which is more likely to be so large that the benefit of a fourth-order scaling does not outweigh the implementation cost that a cyclic-10 design incurs.
Therefore, for the remaining of this paper, we choose cyclic-2 combined with simultaneous diagonalization with cell optimization to be our final proposal, and we no longer consider the cyclic-10 technique as a viable candidate.

\subsection{Molecular Hamiltonian}

Unlike the previous application, the second Hamiltonian we studied in our work has a physical meaning.
That is, we considered the molecular Hamiltonian for dihydrogen.
The corresponding Hamiltonian is computed by leveraging the OpenFermion package \cite{open-fermion}.
\begin{figure*}
    \centering
    \subfigure[]{\includegraphics[width=0.45\linewidth]{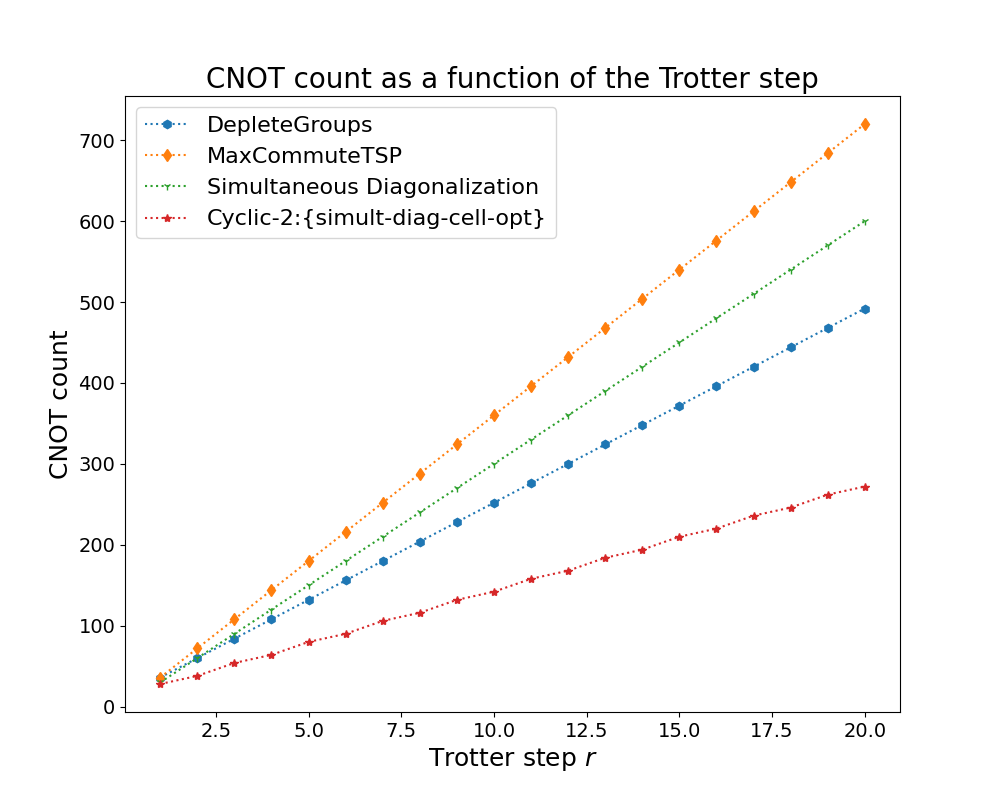} \label{fig:dihydrogen-cnot}}
    \subfigure[]{\includegraphics[width=0.45\linewidth]{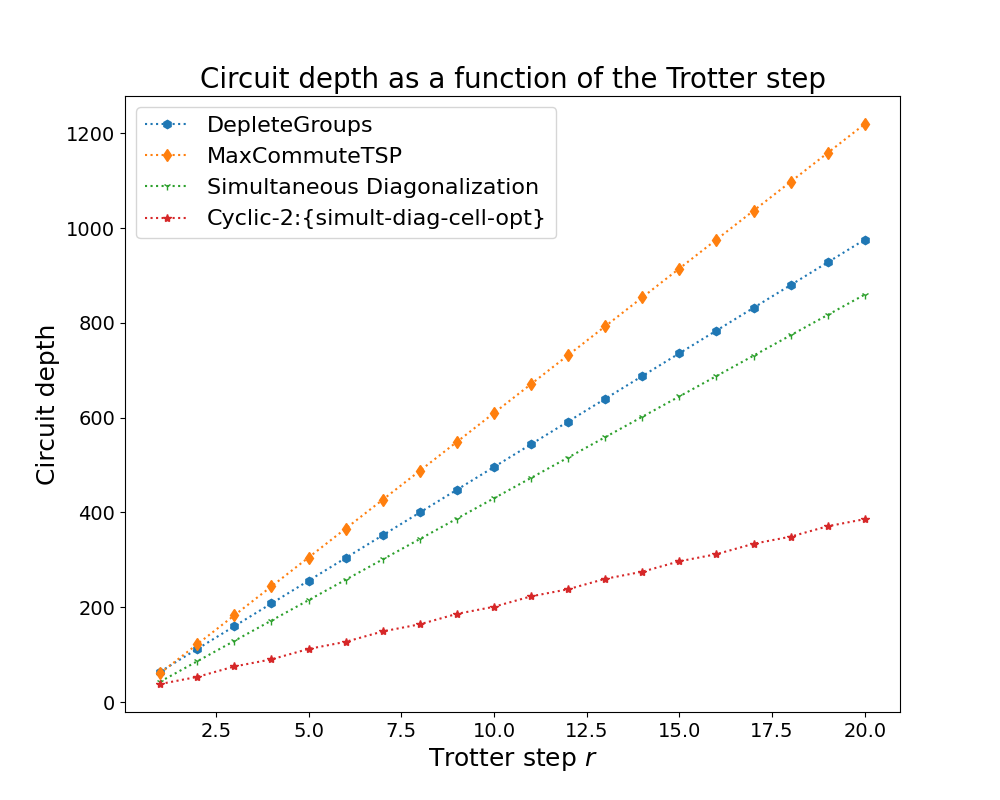} \label{fig:dihydrogen-depth}}
    \subfigure[]{\includegraphics[width=0.45\linewidth]{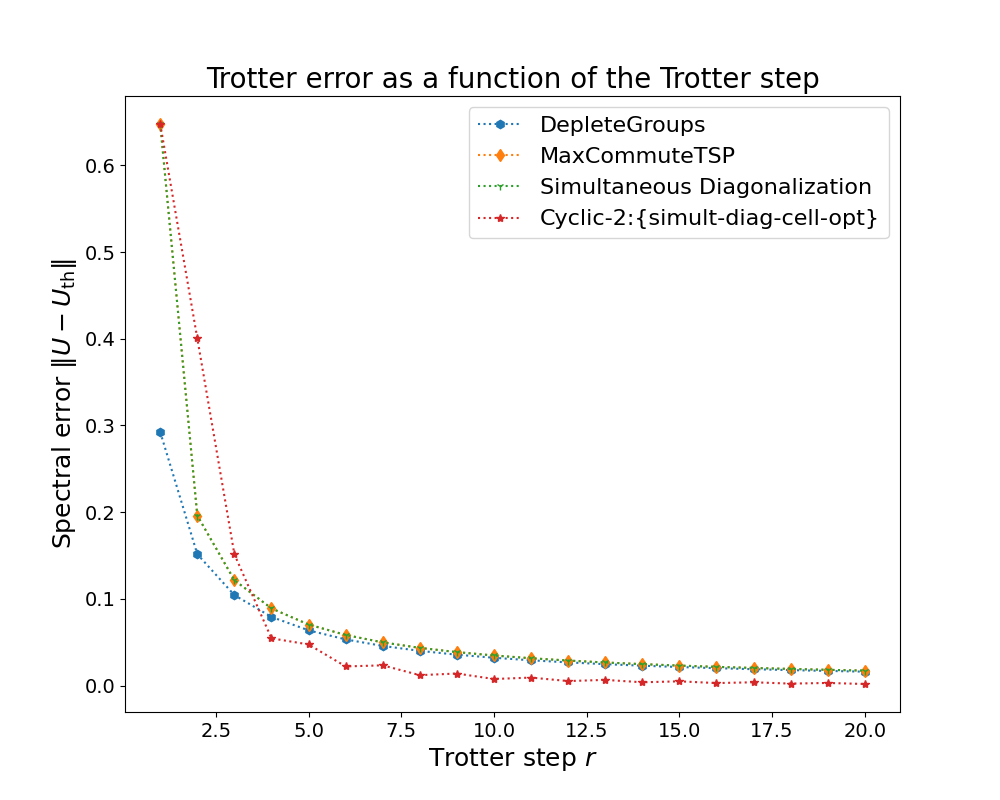} \label{fig:dihydrogen-err}}
    \subfigure[]{\includegraphics[width=0.45\linewidth]{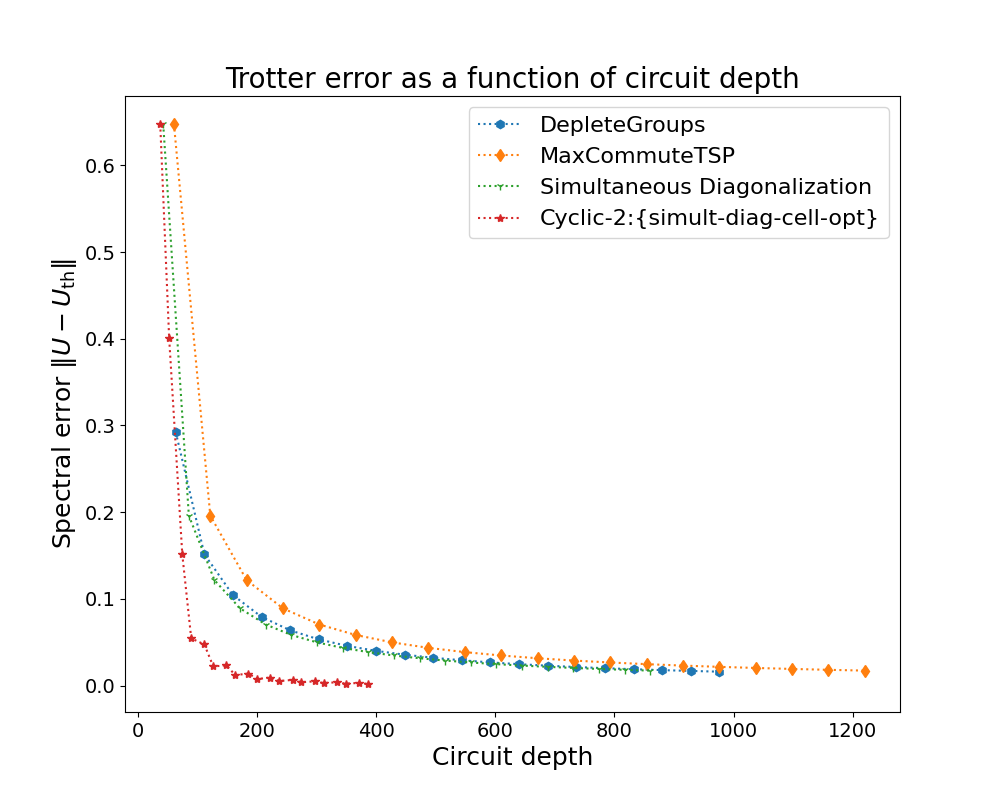} \label{fig:dihydrogen-err-circ}}
    \caption{Performance comparison for the Dihydrogen molecule}
    \label{fig:results-dihydrogen}
\end{figure*}
More precisely, we set the configuration of the molecule as follows
\begin{verbatim}
    geometry = [[`H', [0, 0, 0]],
                [`H', [0, 0, 0.74]]]
    basis = `sto-3g'
    multiplicity = 1
    charge = 0
\end{verbatim}
With the qubit-mapping carried out by the OpenFermion package, we obtained a 4-qubit Hamiltonian system:
\begin{equation}
    \begin{split}
        \mathcal{H} = -0.09707 \, \texttt{IIII} - 0.22343 \, \texttt{IIIZ} - 0.22343 \, \texttt{IIZI} \\
        + 0.17441 \texttt{IIZZ} + 0.17141 \, \texttt{IZII} + 0.12063 \, \texttt{IZIZ} \\
        0.16593 \, \texttt{IZZI} - 0.04530 \, \texttt{XXYY} + 0.04530 \, \texttt{XYYX}\\
        + 0.04530 \, \texttt{YXXY} - 0.04530 \, \texttt{YYXX} + 0.17141 \, \texttt{ZIII} \\
        + 0.16593 \, \texttt{ZIIZ} + 0.12063 \, \texttt{ZIZI} + 0.16869 \, \texttt{ZZII}
    \end{split}
    \label{eq:dihydrogen}
\end{equation}
We ran the experiment with simulation time $t = 3$.
The CNOT count and circuit depth scalings are displayed in Figs.~\ref{fig:dihydrogen-cnot} and \ref{fig:dihydrogen-depth}, respectively.
For the same reasons as detailed earlier with the previous application, cyclic-2 offers clear improvements over previous techniques in terms of the circuit complexity.

The error reduction is presented in Fig.~\ref{fig:dihydrogen-err}.
We can see that for $r \leq 3$, our scheduling strategy performs the worst, and the most attenuated error is achieved by the DepleteGroups strategy.
This is consistent with the findings in \cite{trotter-ordering}, as DepleteGroups proved to offer a considerable improvement over other static arrangements regarding the error reduction in the case of molecular applications.
However for $r > 3$, our proposal surpasses every other strategy including DepleteGroups.
This reveals a peculiar convergence speed which is the expression of the second-order error scaling.

As for the error scaling against the circuit depth (Fig.~\ref{fig:dihydrogen-err-circ}), the improvement is further accentuated thanks to the simultaneous diagonalization with cell optimization mechanism.
From the graph showing the results, we can for example read that to reach the precision achieved with our scheduling strategy for a circuit depth of about 200, all the other strategies require a circuit depth exceeding 1000.

\subsection{Transverse-field Ising model}

Let us move on to another chemistry application.
Here, we consider a spin system that can be described with the transverse-field Ising model:
\begin{equation}
    \mathcal{H} = -J \left(\sum_{q=0}^{n-1} \texttt{Z[q]} \, \texttt{Z[q+1 mod n]} + g \sum_{q=0}^{n-1} \texttt{X[q]}\right),
    \label{eq:ising-model}
\end{equation}
where $n$ designates the number of qubits, $J$ quantifies the interaction between two neighboring sites in the lattice and $g$ represents the coupling with the external magnetic field.
In this scenario, we set $n = 5$, $J = 1$ and $g = 2$.
The simulation time is taken to be $t = 1$.

\begin{figure*}
    \centering
    \subfigure[]{\includegraphics[width=0.45\linewidth]{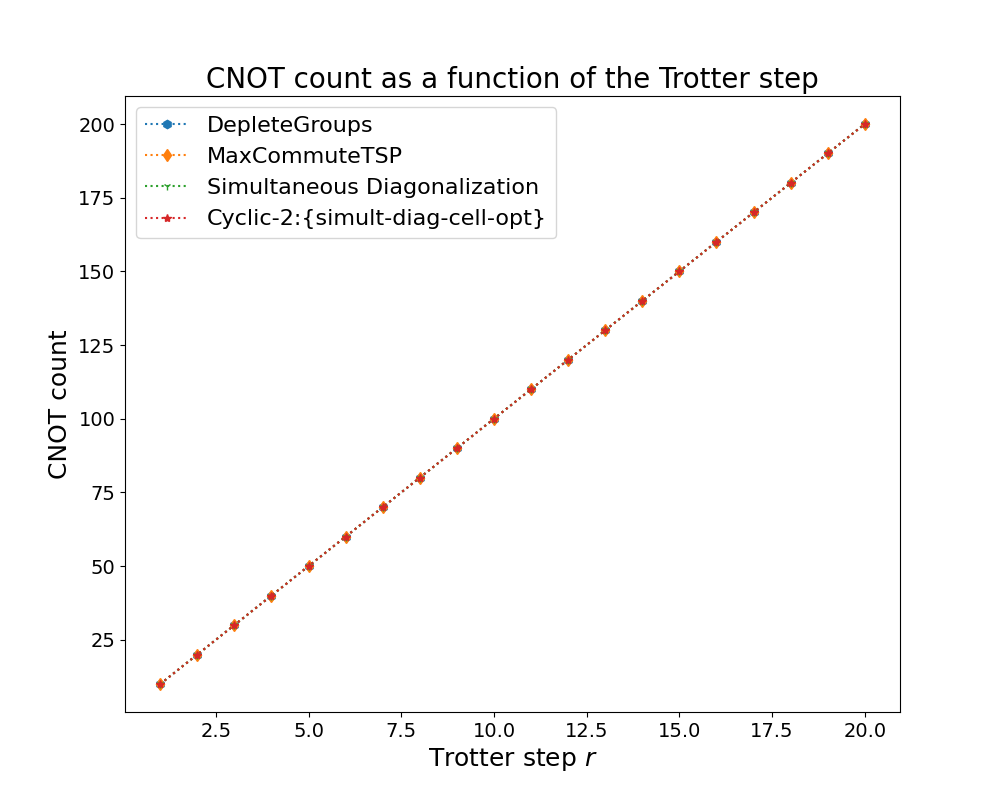} \label{fig:ising-model-cnot}}
    \subfigure[]{\includegraphics[width=0.45\linewidth]{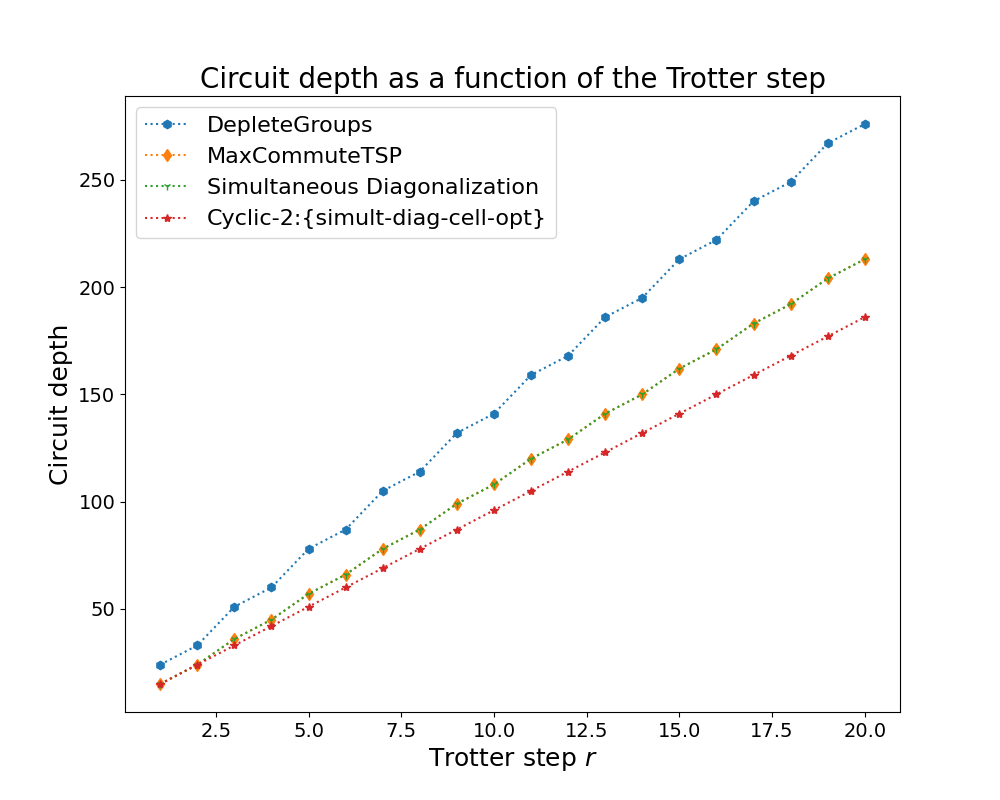} \label{fig:ising-model-depth}}
    \subfigure[]{\includegraphics[width=0.45\linewidth]{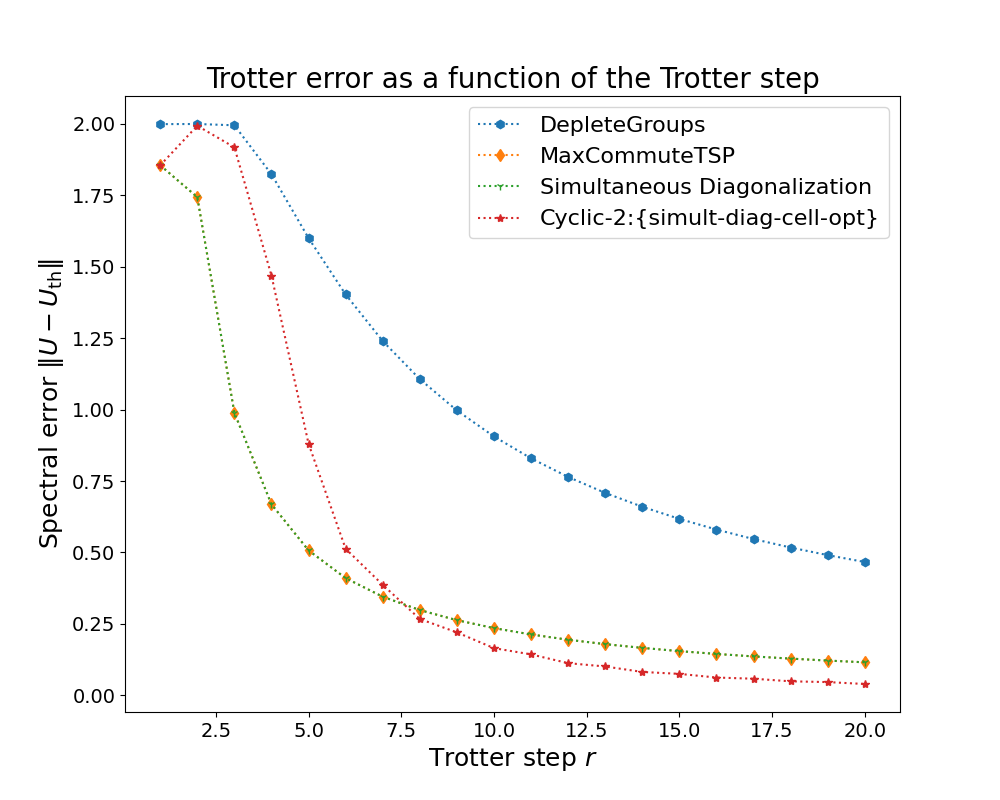} \label{fig:ising-model-err}}
    \subfigure[]{\includegraphics[width=0.45\linewidth]{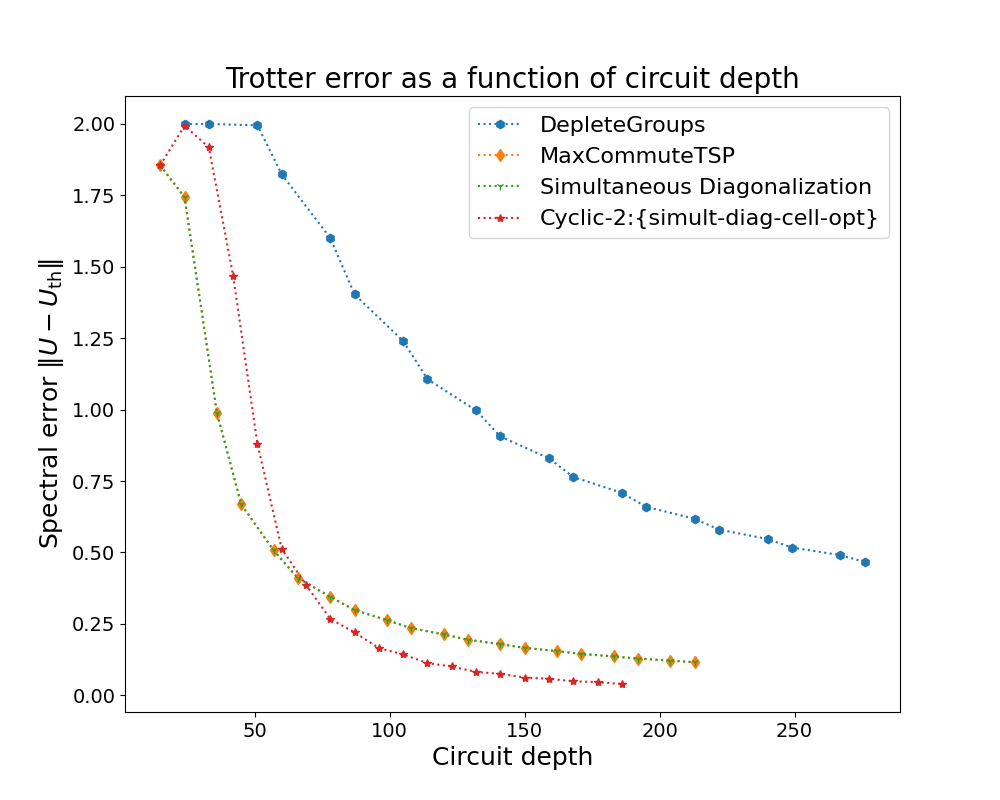} \label{fig:ising-model-err-circ}}
    \caption{Performance comparison for the Ising model ($n=5$, $J=1$, $g=2$)}
    \label{fig:results-ising-model}
\end{figure*}

The results for the CNOT count are given in Fig.~\ref{fig:ising-model-cnot}, showing no difference between the scheduling strategies.
We suspect this is due to the lack of possible CNOT simplification within the commuting subsets found by our graph coloring problem solver.

As for the circuit depth (Fig.~\ref{fig:ising-model-depth}), an improvement can be observed in favor of our strategy similarly to the first two applications.
Interestingly, despite the lack of CNOT simplification, the MaxCommuteTSP technique produces a circuit as shallow as the one based on the traditional version of simultaneous diagonalization.
We conclude that in this situation, the circuit depth cannot be reduced by the simultaneous diagonalization mechanism, and hence there is no advantage in performing it.
However, our proposal provides slightly shorter depths.
Because this improvement occurs while neither CNOT gate has been simplified by the CNOT optimization procedure within each commuting subset nor the simultaneous diagonalization has led to a shallower circuits, it can only be explained by gate simplifications that happen at every transition between forward and backward segments.
Indeed, thanks to the symmetric structure of the cyclic-2 scheme, the last commuting subset of a forward segment is the same as the first commuting subset of a backward segment, so the uncompute-diagonalization stage of the former followed by the compute-diagonalization stage of the latter reduces to the identity operation.
The same naturally applies to a backward-forward transition too.

Disparities in the error reduction appear in Fig.~\ref{fig:ising-model-err}.
It is noteworthy that while we recall a rather efficient performance of the DepleteGroups strategy in terms of the error scaling for molecular systems, it performs very poorly for the Ising model, which confirms its very constrained scope.
Such limitation does not seem to apply to our proposal, as the second-order error scaling (Fig.~\ref{fig:ising-model-err}) gives such an exorbitant advantage for our scheduling strategy in comparison with any other static technique.
The results in the practical framework are shown in Fig.~\ref{fig:ising-model-err-circ}, in which our method appears once again to be the most effective.

\subsection{Heisenberg model}

Lastly, we consider the Heisenberg model:
\begin{equation}
    \begin{split}
        \mathcal{H} = -\frac{1}{2} \sum_{q=0}^{n-1} \Large( & J_x \, \texttt{X[q]} \, \texttt{X[q+1 mod n]} \\
        & + J_y \, \texttt{Y[q]} \, \texttt{Y[q+1 mod n]} \\
        & + J_z \, \texttt{Z[q]} \, \texttt{Z[q+1 mod n]} + h \, \texttt{Z[q]} \Large),
    \end{split}
    \label{eq:heisenberg-model}
\end{equation}
where $J_x$, $J_y$ and $J_z$ correspond to interactions between two neighboring sites in the lattice in the $x$, $y$ and $z$ directions respectively, and $h$ represents the external magnetic field (assumed to be along the $z$ axis).
In our experiment, we set $J_x = J_y = J_z = h = 1$ for a 8-qubit model, with a simulation time $t = 1$.
\begin{figure*}
    \centering
    \subfigure[]{\includegraphics[width=0.45\linewidth]{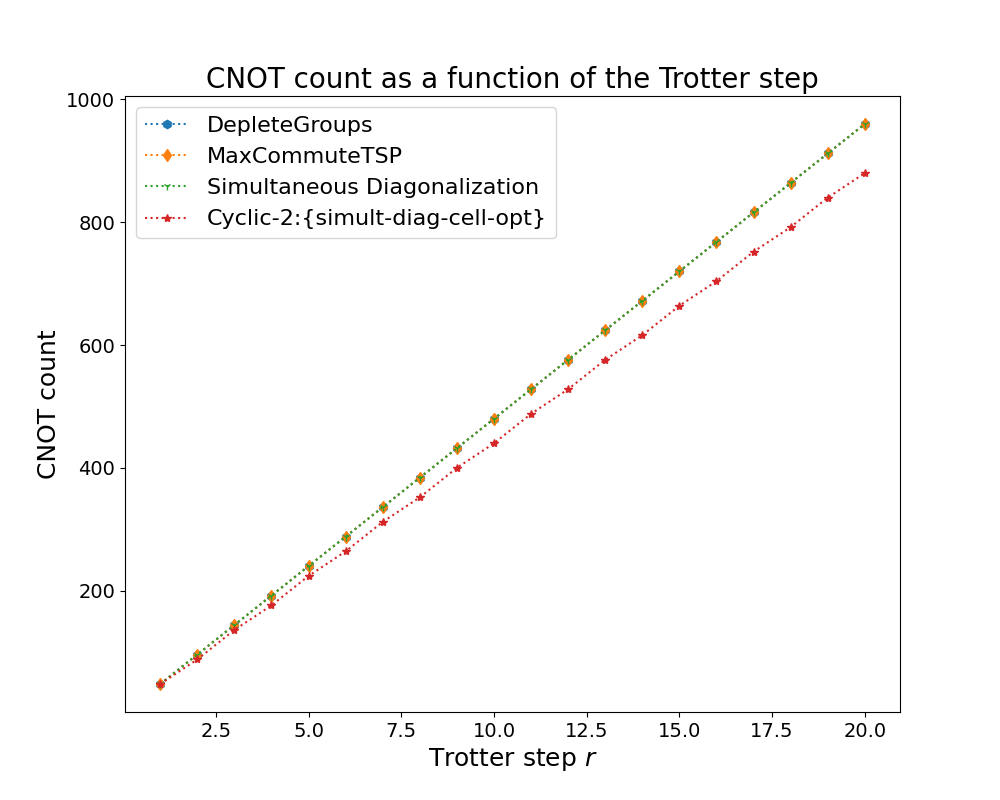} \label{fig:heisenberg-model-cnot}}
    \subfigure[]{\includegraphics[width=0.45\linewidth]{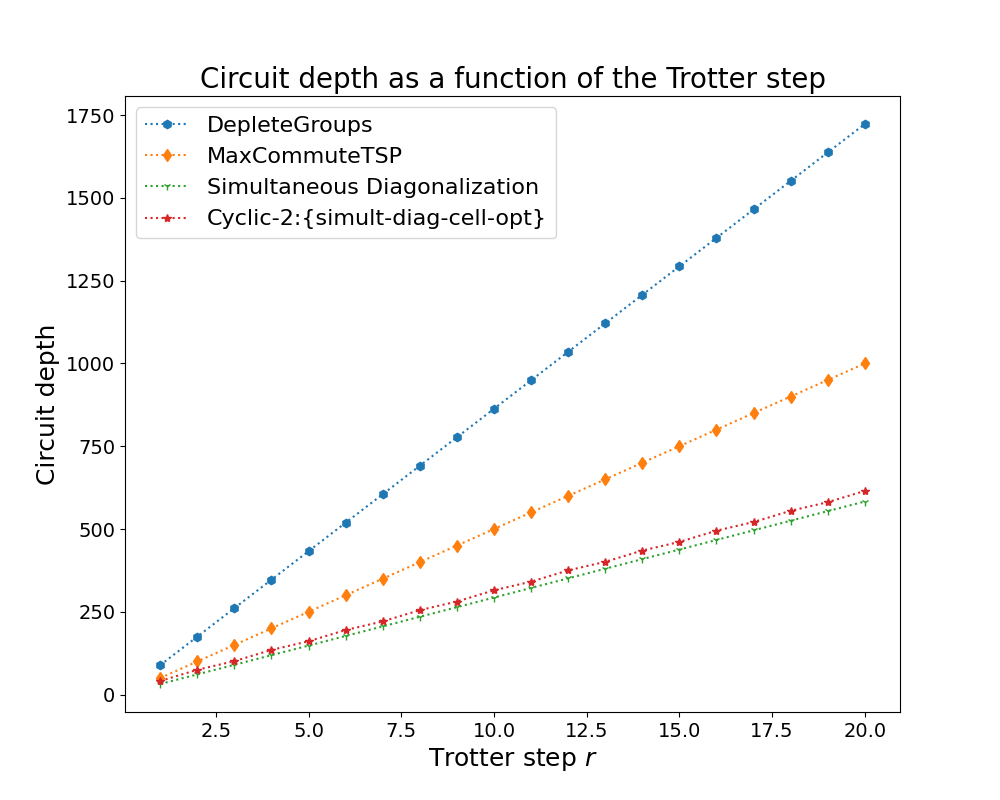} \label{fig:heisenberg-model-depth}}
    \subfigure[]{\includegraphics[width=0.45\linewidth]{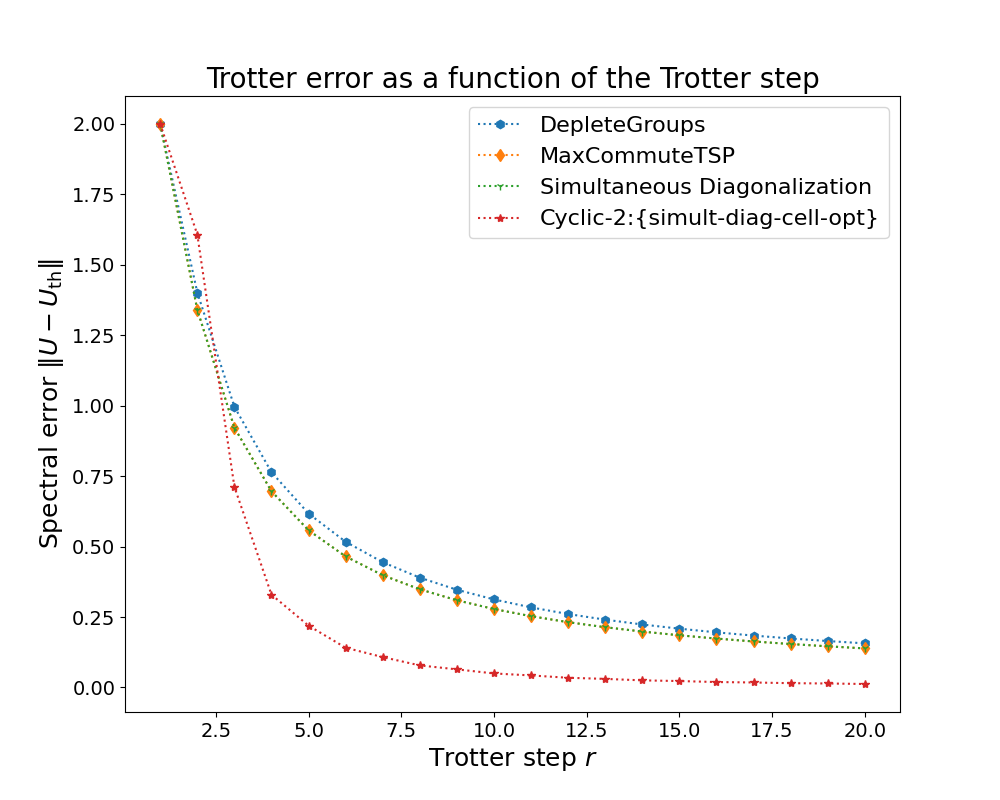} \label{fig:heisenberg-model-err}}
    \subfigure[]{\includegraphics[width=0.45\linewidth]{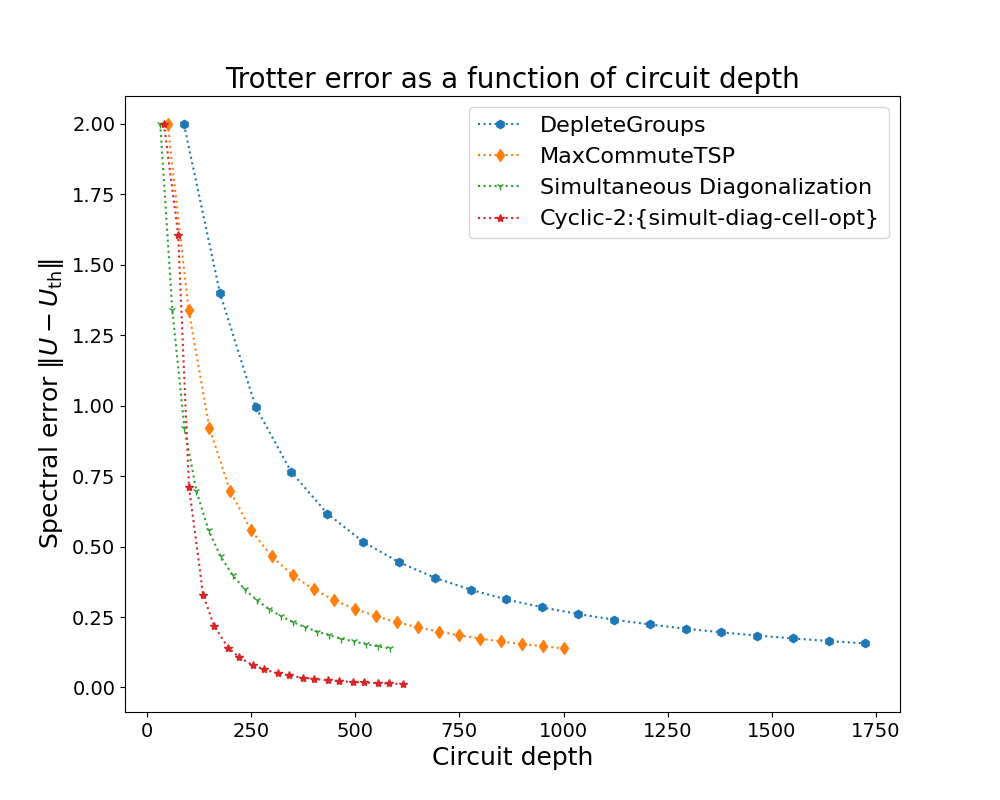} \label{fig:heisenberg-model-err-circ}}
    \caption{Performance comparison for the Heisenberg model ($n=8$, $J_x=1$, $J_y=1$, $J_z=1$, $h=1$)}
    \label{fig:results-heisenberg-model}
\end{figure*}

From Fig.~\ref{fig:heisenberg-model-cnot}, it can be said that the CNOT optimization procedure leads to a slight advantage for our proposal.
As for circuit depths however, Fig.~\ref{fig:heisenberg-model-depth} indicates large disparities between the techniques.
Surprisingly, the traditional simultaneous diagonalization approach seems to outperform our strategy.
Although the difference is very small, such a result is unexpected, as it arises despite all the arguments we gave when we elaborated on the other Hamiltonian systems (i.e. smaller CNOT counts and gate simplifications in forward-backward and backward-forward transitions), suggesting that our proposal should \textit{always} yield shorter quantum circuits.
We believe this is due to a modification of the arrangement within one or several commuting subsets caused by the CNOT optimization procedure, which may prevent other gate simplifications to occur, especially involving the diagonalization stage.
That said, the advantage of the second-order error scaling is so overwhelming (Fig.~\ref{fig:heisenberg-model-err}) that overall, our proposal still stands as the most effective scheduling strategy for this instance of Hamiltonian simulation (Fig.~\ref{fig:heisenberg-model-err-circ}).

\section{Conclusion}
\label{sec:conclusion}

In this work, we proposed some scheduling strategies for the implementation of Hamiltonian simulation via product formulas on quantum computers.
We numerically found that a particularly useful scheme is ``cyclic-2" that uses simultaneous diagonalization with cell optimization as the seed arrangement.
Indeed, this cyclic-2 strategy benefits from both advantages of yielding drastically reduced circuit depths and increasing the convergence speed of the Trotter error compared to previous static methods.
We also presented the extension of our approach corresponding to the fourth-order Trotter-Suzuki decomposition by introducing the ``cyclic-10" mechanism.
From our numerical experiments, however, it seemed that the benefit of a fourth-order scaling can only be reached with a very large number of Trotter steps since it did not prove a notable advantage compared to the second-order technique for $r \leq 20$.
Our view is that the cyclic-2 strategy would achieve a satisfactory performance without the need of prior tests on the Hamiltonian of interest, and this regularity is probably the most appealing contribution of this paper.

Note that, in this work, all the simulations were performed on classical computers, and the complexities of these simulations scale exponentially with the number of qubits, making it impractical to carry out experiments for Hamiltonian systems of bigger sizes.
Thus, further investigations that deal with large numbers of qubits are needed, which will inevitably require to use real quantum hardware.
We expect this to provide opportunities to study our scheduling strategy in a more physical context, where the time spent and the energy consumed by the device can directly be regarded as relevant metrics instead of the circuit depth, against which the actual error is assessed.
Ultimately, the improvements offered by our method are supposed to reflect on the performance of quantum algorithms with real-world applications that leverage Hamiltonian simulation as a subroutine such as the Quantum Eigenvalue Transformation of unitary matrices (QETU) for ground-state preparation and ground-energy estimation \cite{qetu}.

\emph{Note added}: While preparing this manuscript, a preprint by Z. Liu \textit{et al.} appeared on arXiv \cite{new-preprint}, which addresses a similar problem using a different approach.

\begin{acknowledgments}
This work was supported by MEXT Quantum Leap Flagship Program Grants No. JPMXS0118067285 and No. JPMXS0120319794. 
\end{acknowledgments}

\bibliography{refs}

\end{document}